\documentclass[fleqn,12pt,a4paper]{article}
\usepackage{epsf}
\usepackage{sw20lart}
\usepackage{epsfig}
\usepackage{soul}
\usepackage{amsmath}
\usepackage{amssymb}
\usepackage{amsbsy}
\usepackage{amsfonts}
\usepackage{graphics}
\usepackage{latexsym}
\usepackage{cite}
\usepackage[mathscr]{eucal}                      

\newcommand{\ts}{\tau}                           
\usepackage{graphicx}

\newcommand{\bc}{\begin{center}}
\newcommand{\ec}{\end{center}}
\newcommand{\be}{\begin{equation}}
\newcommand{\ee}{\end{equation}}
\newcommand{\bea}{\begin{eqnarray}}
\newcommand{\eea}{\end{eqnarray}}
\newcommand{\bi}{\begin{itemize}}
\newcommand{\ei}{\end{itemize}}
\newcommand{\bt}{\begin{tabular}}
\newcommand{\et}{\end{tabular}}

\def\Nf{n_f}

\def\msbar{\overline{\rm MS}}

\newcommand{\Dlr}{\overset{\leftrightarrow}{D}}

\begin{document}


\date{}
\title{
\begin{flushleft}
\vspace*{-2cm}
{\normalsize DESY 05-107}\\[-0.3em]
{\normalsize Edinburgh 2005/05}\\[-0.3em]
{\normalsize MPP-2005-60}
\vspace*{0.15cm}
\end{flushleft}
Quark helicity flip generalized parton distributions from
two-flavor lattice QCD%
}

\author{
\normalsize M. G\"ockeler$^{1,2}$, Ph. H\"agler$^3$,
R. Horsley$^4$, D. Pleiter$^5$, P.E.L. Rakow$^6$, A. Sch\"afer$^3$,\\
\normalsize G. Schierholz$^{5,7}$, J.M. Zanotti$^5$\\
\hspace{0.5cm}
\normalsize$^{1}$Max-Planck-Institut f\"ur Physik, F\"ohringer Ring 6,
             80805 M\"unchen, Germany \\
\normalsize$^{2}$Institut f\"ur Theoretische Physik, Universit\"at
             Regensburg, 93040 Regensburg, Germany\\
\normalsize$^{3}$Department of Physics and Astronomy, Vrije
             Universiteit, 1081 HV Amsterdam, NL\\
\normalsize$^{4}$School of Physics, University of Edinburgh, Edinburgh
             EH9 3JZ, UK\\ 
\normalsize$^{5}$John von Neumann-Institut f\"ur Computing NIC / DESY,
             15738 Zeuthen, Germany\\
\normalsize$^{6}$Theoretical Physics Division, Dep.~of Math.~Sciences,
             University of Liverpool, \\ 
\normalsize  Liverpool L69 3BX, UK\\
\normalsize$^{7}$Deutsches Elektronen-Synchrotron DESY, 22603 Hamburg,
             Germany\\ 
\normalsize\emph{(QCDSF/UKQCD Collaboration)}}

\maketitle

\begin{abstract}
  We present an initiatory study of quark helicity flip generalized
  parton distributions (GPDs)  in
  $n_f=2$ lattice QCD, based on clover-improved Wilson fermions for a
  large number of coupling constants and pion masses.
  Quark helicity flip GPDs yield essential information on the
  transverse spin structure of the nucleon. In this work, we show
  first results on their lowest moments and dipole masses and study
  the corresponding chiral and continuum extrapolations.
\end{abstract}

\vspace*{-2mm}
\section{Introduction}
\vspace*{-2mm}

Generalized parton distributions (GPDs) \cite{GPD} have opened new
ways of studying the complex interplay of longitudinal momentum and
transverse coordinate space \cite{Burkardt:2000za,Diehl:2002he}, as
well as spin and orbital angular momentum degrees of freedom in the
nucleon \cite{Ji:1996ek}.
\begin{figure}[t]
\bc
\includegraphics[height=10cm,angle=-90]{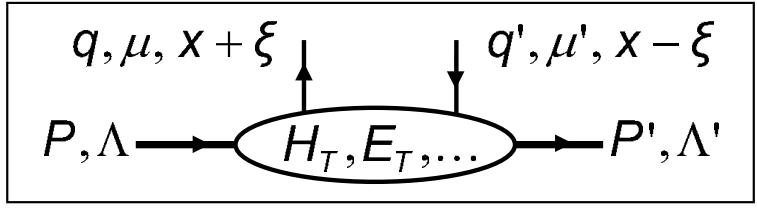}
\caption{The lower part of the handbag diagram.}
\label{fig:Amplitude}
\ec
\end{figure}
As a counting of the helicity amplitudes in Fig.~\ref{fig:Amplitude}
reveals \cite{Diehl:2001pm}, there are eight independent real
functions needed at twist 2.
Four of them, namely $H_T$, $E_T$, $\widetilde H_T$ and $\widetilde
E_T$, are related to a flip of the quark helicity, $\mu=-\mu'$, hence
{\it quark helicity flip} GPDs\footnote{Also called tensor GPDs.}.
Quark helicity flip GPDs play a prominent role in the understanding of
the transverse spin structure of the nucleon and significantly sharpen
positivity bounds on GPDs in impact parameter space
\cite{Diehl:2005jf}.
Specifically, it could be very interesting to exploit and study the
equation-of-motion relations between the lowest moments of quark
helicity flip, unpolarized and twist-3 GPDs which have been obtained
in \cite{Diehl:2005jf}.
The (chirally odd) tensor GPDs also provide a framework with which to
study the correlation between quark spin and quark angular momentum in
unpolarized nucleons \cite{Burkardt:2005hp}.

Quark helicity flip GPDs are defined via the parameterization of an
off-forward nucleon matrix element of a quark operator involving the
$\sigma^{\mu\nu}$-tensor as follows \cite{Diehl:2001pm}
\begin{eqnarray}
   \lefteqn{\hspace*{-8mm}\left\langle P',\Lambda ' \right|
  \int \frac{d \lambda}{4 \pi} e^{i \lambda x}
  \bar \psi (-\frac{\lambda}{2}n)
  \, i\sigma^{\mu\nu}
  \psi(\frac{\lambda}{2} n)
  \left| P,\Lambda \right\rangle
   =\overline U(P',\Lambda ') \bigg( i\sigma^{\mu\nu} H_T(x, \xi, t)
 + \frac{\gamma^{[\mu}\Delta^{\nu]}} {2 m} E_T(x, \xi, t) 
 \nonumber}\\
& &\mbox{}+  \frac{\overline P^{[\mu} \Delta^{\nu]}} {m^2 } \widetilde H_T(x, \xi, t)
 + \frac{\gamma^{[\mu}\overline P^{\nu]}} {m} \widetilde E_T(x, \xi, t)\!\bigg)   U(P,\Lambda)
 \label{GPDs1}.
\end{eqnarray}
Here the momentum transfer is given by $\Delta=P'-P$ with
$t=\Delta^2$, $\overline P = (P'+P)/2$, and $\xi=-n\cdot \Delta/2$
denotes the longitudinal momentum transfer, where $n$ is a light-like
vector.
The first of these  tensor GPDs, $H_T(x,\xi,t)$, is called
generalized transversity, because it reproduces the transversity
distribution in the forward limit, $H_T(x,0,0)=\delta q(x)=h_1(x)$.
Integrating $H_T(x, \xi, t)$ over $x$ gives the tensor form factor:
\begin{equation}
\int_{-1}^{1} dx H_T(x, \xi, t) = g_T(t).
\end{equation}

Since the quark tensor GPDs require a helicity flip of the quarks,
they do not contribute to the deeply virtual Compton scattering (DVCS)
process $\gamma^* p\rightarrow \gamma p'.$
Naively, one could think that this could be balanced by the production
of a transversely polarized vector meson instead of a photon,
$\gamma^* p\rightarrow m_T p'$.
However, it has been shown that the corresponding amplitude,
remarkably, vanishes at leading twist to all orders in perturbation
theory \cite{Mankiewicz:1997uy,Diehl:1998pd,Collins:1999un}.
The only process giving access to the generalized transversity which
has been proposed in the literature so far is the diffractive double
meson production $\gamma^* p\rightarrow m_L m_T p'$
\cite{Ivanov:2002jj}.
Naturally, one expects the measurement of this reaction to be much
more involved than e.g.\ the exclusive electroproduction of a single
vector meson.
Since the tensor GPDs are practically unknown, it is unclear how to
even estimate the corresponding cross section to see if a measurement
of this process is at all feasible.
Given that the situation seems to be much more difficult than for the
(un-)polarized GPDs, lattice calculations of the lowest moments of the quark
helicity flip GPDs will be highly valuable. While (un-)polarized GPDs
have already been investigated in a number of papers
\cite{QCDSF-1,Gockeler:2004vx,Gockeler:2004mn,Gockeler:2005aw,MIT,MIT-2,Hagler:2004er,Renner:2005sm},
we present here the first lattice calculation of quark helicity flip GPDs.

Lattice calculations of moments of parton distributions mostly
disregard the computationally expensive quark-line disconnected
contributions.
They correspond to a situation where the operator is inserted into a
closed quark loop which is connected to the nucleon only via gluons.
Since the tensor operators flip the quark helicity, these disconnected
diagrams do not contribute in the continuum theory for vanishing quark
masses. Therefore, we expect only small contributions for the 
disconnected graphs in our calculation.
This expectation is supported by numerical results from
\cite{Aoki:1996pi}, where the tensor charge was calculated in quenched lattice QCD.
The authors explicitly computed the disconnected pieces for the tensor
operator and found the contributions from up- and down-quarks to be
compatible with zero within one standard deviation.
Thus it is possible to estimate the individual up and down quark
tensor GPDs, which is a major advantage compared to other
observables where usually only the iso-vector channel is considered.
Further early results on the tensor charge in quenched lattice QCD
have been presented in \cite{Gockeler:1996es,Capitani:1999zd}.

As mentioned above, in calculating the lowest moments of the tensor
GPD $H_T(x,\xi,t)$, we automatically obtain the corresponding moments
of the transversity distribution, $\langle x^{n-1} \rangle_\delta$,
for $t=\xi=0$.
The quark transversity has recently attracted renewed attention
related to the Collins asymmetry in e.g.~semi-inclusive deep inelastic
scattering.
It is generally believed that transverse single-spin asymmetries (SSA)
\cite{Mulders:1995dh} are generated predominantly by the Sivers and
Collins mechanism.
These two differ in their dependence on the azimuthal angles and thus
can be separated.  The contribution due to the Collins mechanism is
proportional to a convolution of the transversity distribution $\delta
q(x)$ and the Collins fragmentation function $H_1^\perp(z)$, which are
both chiral odd.
Lack of knowledge of both the transversity and the Collins function,
however, seriously hampers the interpretation of the exciting
experimental results on such SSAs
\cite{Airapetian:2004tw,Alexakhin:2005iw}.
Lattice results for the lowest moments of $\delta q(x)$ for up and
down quarks could help to reveal the physics behind these measured
asymmetries.

The paper is organized as follows.
We begin by briefly reminding the reader of the methods and techniques
we use to extract moments of GPDs from the lattice in Section~2.
In Section~3, we specify the parameters of our calculation and present
our results for the lowest moments of the tensor GPD $H_T(x,\xi,t)$.
Making use of the large number of results for different sets of
lattice parameters, we attempt to extrapolate the moments of the
generalized transversity as well as the dipole masses of the tensor
GPDs to the continuum and chiral limits.
Finally, in Section~4 we summarize our findings.

\vspace*{-3mm}
\section{Extracting moments of GPDs from lattice simulations}
\vspace*{-2mm}
\label{moments}

On the lattice, it is not possible to deal directly with matrix
elements of bi-local light-cone operators.
Therefore, we first transform the LHS of Eq.~(\ref{GPDs1}) to Mellin
space by integrating over $x$, i.e. $\int_{-1}^{1} dx x^{n-1}$.
This results in nucleon matrix elements of towers of local tensor
operators
\begin{equation}
{\cal O}_T^{\mu\nu \mu_{1}\ldots \mu _{n-1}}(0)= \bar{\psi}(0)
i\sigma^{\mu \{\nu }i\Dlr{}^{\mu_{1}}\ldots
i\Dlr{}^{\mu_{n-1}\}}\psi(0)\, ,
\label{eq:ops}
\end{equation}
which are in turn parameterized in terms of the tensor generalized
form factors (GFFs) $A_{Tni}$, $B_{Tni}$, $\widetilde A_{Tni}$ and
$\widetilde B_{Tni}$.
Here and in the following, $\Dlr = \frac{1}{2}(\overrightarrow{D} -
\overleftarrow{D})$ and $\{\cdots\}$ indicates symmetrization of
indices and subtraction of traces.
The parameterization for arbitrary $n$ is given in
\cite{Hagler:2004yt,Chen:2004cg}
\footnote{Note that the Mellin-moment index $n$ used here differs from
  the number of covariant derivatives $n$ in \cite{Hagler:2004yt} by
  one.}.
Here we show explicitly only the expressions for the lowest two
moments.  For $n=1$ we have
\begin{eqnarray}
\left\langle P^{\prime}\Lambda^{\prime}\right| 
 \bar{\psi}(0) i\sigma^{\mu\nu} \psi(0)
\left| P \Lambda\right\rangle
&=&\overline{U}(P^{\prime},\Lambda^{\prime}) 
 \left\{ i\sigma^{\mu\nu} A_{T10}(t) \right.
+\frac{\overline{P}^{[\mu}\Delta^{\nu ]}} 
      {m^{2}}\widetilde{A}_{T10}(t) \nonumber \\
&+&\left. \frac{\gamma^{[\mu}\Delta^{\nu ]}}{2m}
  B_{T10}(t)\right\} U(P,\Lambda) \ .  
\label{tn0}
\end{eqnarray}
The inclusion of an additional term $\varpropto \gamma^{[\mu
}\overline{P}^{\nu ]}\equiv \gamma^{\mu }\overline{P}^{\nu}-
\gamma^{\nu}\overline{P}^{\mu}$ in Eq.~(\ref{tn0}) is forbidden by
time reversal symmetry \cite{Diehl:2001pm}.
For $n=2$, however, this can be balanced by including another factor
of $\Delta $, leading to four generalized form factors,
\begin{eqnarray}
A_{[\mu \nu ]}S_{\{\nu \mu _{1}\}}
\left\langle P^{\prime}\Lambda^{\prime}\right| 
 \bar{\psi}(0) i\sigma^{\mu\nu} i\Dlr{}^{\mu_{1}}\psi(0) 
\left| P \Lambda\right\rangle 
\!\!&=&\!\!
A_{[\mu\nu]} S_{\{\nu\mu_{1}\}} \overline{U}(P^{\prime},\,\Lambda^{\prime})
\left\{ i\sigma^{\mu \nu}%
\overline{P}^{\mu _{1}}A_{T20}(t)\right.  \nonumber \\
\!\!&+&\!\!\frac{\overline{P}^{[\mu}\Delta^{\nu]}}{m^{2}}
\overline{P}^{\mu_{1}} \widetilde{A}_{T20}(t)
+\frac{\gamma^{[\mu}\Delta^{\nu]}}{2m} \overline{P}^{\mu_{1}}
B_{T20}(t)  \nonumber \\
\!\!&+&\!\!\left. \frac{\gamma^{[\mu} \overline{P}^{\nu]}}{m} \Delta^{\mu_{1}}
\widetilde{B}_{T21}(t)\right\} U(P,\Lambda), \label{tn1}
\end{eqnarray}
up to trace terms, where $A_{[\mu\nu]}$ and $S_{\{\mu\nu\}}$ denote anti-symmetrization
and symmetrization of $(\mu,\nu)$, respectively.
For $n=3$ there are seven independent tensor GFFs, as an explicit
counting shows~\cite{Hagler:2004yt,Chen:2004cg}. The simultaneous
extraction of such a large number of GFFs poses a challenge for
lattice QCD calculations, which we plan to address in the near future.

Instead of calculating continuum Minkowski space-time matrix elements
(e.g.\ in Eqs.~(\ref{tn0}) and (\ref{tn1})) directly, on the lattice
we work within a discretized Euclidean space-time framework to
calculate nucleon two- and three-point correlation functions.
The nucleon two- and three-point functions are given by
\begin{eqnarray}
  C^{\text{2pt}}(\tau,P) &=& \sum_{j,k}
    \tilde\Gamma_{jk}\left\langle N_{k}
    (\tau,P) \overline{N}_{j}(\tau_{\text{src}},P)\right\rangle,\nonumber \\
  C_{\cal O}^{\text{3pt}\mu\nu \mu_{1}\ldots
    \mu_{n-1}}(\tau,P',P) &=& \sum_{j,k}
    \tilde\Gamma_{jk}\left\langle N_{k}
    (\tau_{\text{snk}},P'){\cal O}_T^{\mu\nu \mu_{1}\ldots \mu _{n-1}}(\tau)
  \overline{N}_{j}(\tau_{\text{src}},P)\right\rangle,
  \label{threept}
\end{eqnarray}
where $\tilde \Gamma$ is a (spin-)projection matrix and the operators
$\overline{N}$ and $N$ create and destroy states with the quantum
numbers of the nucleon, respectively.
The relation of $C_{\cal O}^{\text{3pt}}$ to the parameterizations in
Eqs.~(\ref{tn0}) and (\ref{tn1}) is seen by rewriting Eq.~(\ref{threept})
using complete sets of states and the time
evolution operator,
\begin{eqnarray}
  \label{threept2}
C_{\cal O}^{\text{3pt}\mu\nu \mu_{1}\ldots
    \mu_{n-1}}(\ts,P',P) \!\!\!&=&\!\!\!
  \frac{\left(Z(P)\overline{Z}(P')\right)^{1/2}}{4 E(P')E(P)}
  e^{-E(P)(\ts-\ts_{\text{src}})-E(P')(\ts_{\text{snk}}-\ts)}
   \nonumber\\
    &\times& \!\!\!\sum_{\Lambda,\Lambda^{\prime }}
    \left\langle P',\Lambda^{\prime }\right|{\cal O}_T^{\mu\nu \mu_{1}\ldots \mu _{n-1}}
    \left|P,\Lambda \right\rangle \overline{U}(P,\Lambda)\tilde\Gamma
    U(P',\Lambda')+\ldots\ .
\end{eqnarray}
Similarly, the two-point function for $\tilde\Gamma=1/2 (1+\gamma_4)$ can be written as
\begin{equation}
C^{\text{2pt}}(\tau,P)=\left(Z(P)\overline{Z}(P)\right)^{1/2} \frac{E(P)+m}{E(P)}
e^{-E(P)(\ts-\ts_{\text{src}})} + \ldots\ .
\label{twopt2}
\end{equation}
The ellipsis in Eq.~(\ref{threept2}) and (\ref{twopt2}) represents excited states with energies $E'>E(P),E(P')$,
which are exponentially suppressed as long as
$\ts-\ts_{\text{src}}\gg 1/E', \ts_{\text{snk}}-\ts\gg 1/E'$.
%
Inserting the explicit parameterizations from Eqs.~(\ref{tn0}) and
(\ref{tn1}) transformed to Euclidean space into Eq.~(\ref{threept2}), we sum over polarizations
to obtain
\begin{eqnarray}
  \label{threept3}
\lefteqn{C_{\cal O}^{\text{3pt}\mu\nu \mu_{1}\ldots
    \mu_{n-1}}(\ts,P',P)
  =\frac{\left(
      Z(P)\overline{Z}(P')\right)^{1/2}}{4 E(P')E(P)}
  e^{-E(P)(\ts-\ts_{\text{src}})-E(P')(\ts_{\text{snk}}-\ts)}
   \text{Tr}\bigg[ \tilde\Gamma(i\!\!\!\not{\!P}^{\prime}-m)}\nonumber\\
&&\hspace*{2.8cm}   \times
   \bigg(a_{T}^{\mu\nu \mu_{1}\ldots \mu _{n-1}}A_{Tn0}(t)
+   b_{T}^{\mu\nu \mu_{1}\ldots \mu _{n-1}}B_{Tn0}(t)+\cdots \bigg)
   (i\!\!\!\not{\!P}-m)
  \bigg]\ ,
  \label{trace}
\end{eqnarray}
where e.g.~$a_T^{\mu \nu\mu_1}$ is the Euclidean version of the
prefactor $i\sigma^{\mu\nu} \overline P^{\mu_1}$ in Eq.~(\ref{tn1}).
The Dirac-trace in Eq.~(\ref{threept3}) is evaluated explicitly, while
the normalization factor and the exponentials in Eq.~(\ref{threept2})
are cancelled out by constructing an appropriate $\tau$-independent
ratio $R$ of two- and three-point functions,
\begin{equation}
R_{\cal O}(\tau,P', P)\, =
 \frac{C^{\text{3pt}}_{\cal O} (\tau,P', P)}
        {C^{\text{2pt}}(\tau_{\text{snk}},P')}
 \left[
  \frac{C^{\text{2pt}}(\tau,P') C^{\text{2pt}}(\tau_{\text{snk}},P') 
     C^{\text{2pt}}(\tau_{\text{snk}}-\tau+\tau_{\text{src}},P)}
       {C^{\text{2pt}}(\tau,P) C^{\text{2pt}}(\tau_{\text{snk}},P) 
     C^{\text{2pt}}(\tau_{\text{snk}}-\tau+\tau_{\text{src}},P')}
\right]^{\frac{1}{2} } \ .
\label{eq:ratio}
\end{equation}
The ratio $R$ is evaluated numerically and then equated with the
corresponding sum of GFFs times $P$- and $P'$-dependent calculable
pre-factors, coming from the traces in Eq.~(\ref{threept3}).
For a given moment $n$, this is done simultaneously for all
contributing index combinations $(\mu\nu \mu_{1}\ldots \mu _{n-1})$
and all discrete lattice momenta $P,P'$ corresponding to the same
value of $t=(P'-P)^2$.
This procedure leads, in general, to an overdetermined set of
equations from which we finally extract the GFFs \cite{MIT}.
We have taken care to ensure that our normalization leads exactly to
the $x$-moment of the transversity distribution $\delta q(x)=h_1(x)$
as defined in \cite{Jaffe:1991kp}.
To make this as transparent as possible, we give an explicit example
of one of the equations we use to extract $\langle x\rangle_\delta$
\begin{equation}
R^{2\{34\}}=
\frac{C_{\cal O}^{\text{3pt}\,2\{34\}}\!\left(\ts,P'=(m,\vec{0}),P=(m,\vec{0})\right)}
{C^{\text{2pt}}\!\left(\ts_{\text{snk}},P=(m,\vec{0})\right)}
=\frac{1}{2\kappa}\frac{m}{2}\langle x\rangle_\delta\,,
\end{equation}
where only the $\tilde\Gamma_1$ (see Eq.~(\ref{projectors})) projector
contributes and $2\{34\}$ represents the operator $\bar \psi
\sigma^{2\{3}\Dlr{}^{4\}} \psi$.

On the lattice the space-time symmetry is reduced to the hypercubic
group $H(4)$, and the lattice operators have to be chosen such that
they belong to irreducible multiplets under $H(4)$.
Furthermore, one would like to avoid mixing under renormalization as
far as possible.  In the case of the twist-2 operators in
Eq.~(\ref{eq:ops}), or more precisely their Euclidean counterparts,
this presents no problem for $n=1$ and $n=2$, the only cases to be
considered in this paper.
For $n=1$ we have the 6-dimensional multiplet consisting of the
operators 
\begin{equation} 
\bar{\psi}(0) i \sigma_{\mu \nu} \psi (0)  \,,
\end{equation}
which is irreducible in the continuum as well as on the lattice
($H(4)$ representation $\tau^{(6)}_1$ in the notation of \cite{bonn}).
The 16-dimensional space of continuum twist-2 operators with $n=2$
decomposes into two 8-dimensional multiplets transforming according to
the inequivalent representations $\tau^{(8)}_1$ and $\tau^{(8)}_2$.
Typical members of these multiplets are, e.g.,
\begin{equation}
 \bar{\psi}(0) \left( i \sigma_{12} \overset{\leftrightarrow}{D}_2
  - i \sigma_{13} \overset{\leftrightarrow}{D}_3 \right) \psi (0) 
\end{equation}
in the case of $\tau^{(8)}_1$, and
\begin{equation}
 \bar{\psi}(0) \left( i \sigma_{12} \overset{\leftrightarrow}{D}_3
  + i \sigma_{13} \overset{\leftrightarrow}{D}_2 \right) \psi (0)
\end{equation} for $\tau^{(8)}_2$.
All these operators are free of mixing problems, but one has to take
into account that operators belonging to inequivalent representations
have different renormalization factors.

Obviously, for a successful computation of the GFFs, one would like to
have as many different nucleon sink and source momenta and projection
operators $\tilde\Gamma$ as possible in order to obtain a large number
of independent non-vanishing Dirac-traces in Eq.~(\ref{trace}).
This is particularly true for the tensor operators because they
involve $\sigma^{\mu\nu}$ and the number of tensor GFFs grows rapidly
with $n$.
Once we have extracted the GFFs from the lattice correlation
functions, it is an easy exercise to reconstruct the corresponding
moments of tensor GPDs, $H^n_{T}(\xi ,t)=\int dx x^{n-1}H_T(x,\xi,t)$
etc., using the polynomiality relations \cite{Hagler:2004yt}
\begin{equation}
\begin{array}{lcl}
H^{n=1}_{T}(\xi,t) = A_{T10}(t) = g_T(t), & \qquad &
  H^{n=2}_{T}(\xi,t) = A_{T20}(t), \\
\widetilde{H}^{n=1}_{T}(\xi,t) = \widetilde{A}_{T10}(t), & \qquad &
  \widetilde{H}^{n=2}_{T}(\xi,t) = \widetilde{A}_{T20}(t), \\
E^{n=1}_{T}(\xi,t) = B_{T10}(t), & \qquad &
  E^{n=2}_{T}(\xi,t) = B_{T20}(t), \\
\widetilde{E}^{n=1}_{T}(\xi,t) = \widetilde{B}_{T10}(t) = 0, & \qquad &
  \widetilde{E}^{n=2}_{T}(\xi,t) = (-2\xi)\widetilde{B}_{T21}(t)\,.
\label{poly}
\end{array}
\end{equation}
These equations directly show that for $n\leq 2$, a dependence on the
longitudinal momentum transfer $\xi$ is only seen for the GPD
$\widetilde{E}_{T}$, which is the only quark GPD odd in $\xi$.
\begin{table}[tb]
\begin{center}
  \begin{tabular}{cccccc}
        $\beta$ & $\kappa_{\rm sea}$ & Volume & $N_{\rm traj}$ & $a$~(fm) &
        $m_{\pi}$~(GeV)     \\ \hline
    5.20 & 0.13420 & $16^3\times 32$ & ${\cal O}$(5000) & 0.1145 & 1.007(2)    \\
    5.20 & 0.13500 & $16^3\times 32$ & ${\cal O}$(8000) & 0.0982 & 0.833(3)    \\
    5.20 & 0.13550 & $16^3\times 32$ & ${\cal O}$(8000) & 0.0926 & 0.619(3)    \\
    5.25 & 0.13460 & $16^3\times 32$ & ${\cal O}$(5800) & 0.0986 & 0.987(2)    \\
    5.25 & 0.13520 & $16^3\times 32$ & ${\cal O}$(8000) & 0.0909 & 0.829(3)    \\
    5.25 & 0.13575 & $24^3\times 48$ & ${\cal O}$(5900) & 0.0844 & 0.597(1)   \\
    5.29 & 0.13400 & $16^3\times 32$ & ${\cal O}$(4000) & 0.0970 & 1.173(2)    \\
    5.29 & 0.13500 & $16^3\times 32$ & ${\cal O}$(5600) & 0.0893 & 0.929(2)   \\
    5.29 & 0.13550 & $24^3\times 48$ & ${\cal O}$(2000) & 0.0839 & 0.769(2)   \\
    5.40 & 0.13500 & $24^3\times 48$ & ${\cal O}$(3700) & 0.0767 & 1.037(1)   \\
    5.40 & 0.13560 & $24^3\times 48$ & ${\cal O}$(3500) & 0.0732 & 0.842(2)      \\
    5.40 & 0.13610 & $24^3\times 48$ & ${\cal O}$(3500) & 0.0696 & 0.626(2)
\end{tabular}
\caption{Lattice parameters:
 gauge coupling $\beta$, sea quark hopping parameter $\kappa_{sea}$,
 lattice volume, number of trajectories, lattice spacing and
 pion mass.
  \label{table:parameters}}
\end{center}
\vspace{-0.5cm}
\end{table}
In order to investigate the $\xi$ dependence of the generalized
transversity $H^n_{T}(\xi ,t)$, one has to consider at least the
$n=3$ Mellin moment.
Finally, we note that in the forward limit the moments $H^n_{T}(\xi
,t)$ reduce to the moments of the transversity distribution,
$H^n_{T}(\xi=0 ,t=0)=\langle x^{n-1} \rangle_\delta$.


\vspace*{-3mm}
\section{Lattice results for moments of the generalized \newline transversity}
\vspace*{-2mm}

The simulations are done with $\Nf=2$ flavors of
dynamical non-perturbatively ${\cal O}(a)$ improved Wilson fermions and Wilson glue.
For four different values $\beta=5.20$, $5.25$, $5.29$, $5.40$ and
three different $\kappa$ values per $\beta$ we have in collaboration
with UKQCD generated ${\cal O}(2000-8000)$ trajectories.
Lattice spacings and spatial volumes vary between 0.07-0.11~fm and
(1.4-2.0~fm)$^3$ respectively.
A summary of the parameter space spanned by our dynamical
configurations can be found in Table~\ref{table:parameters}.  
We set the scale via the force parameter $r_0$, with $r_0=0.467$~fm.

\begin{figure}[t]
\bc
\includegraphics[height=15cm,angle=-90]{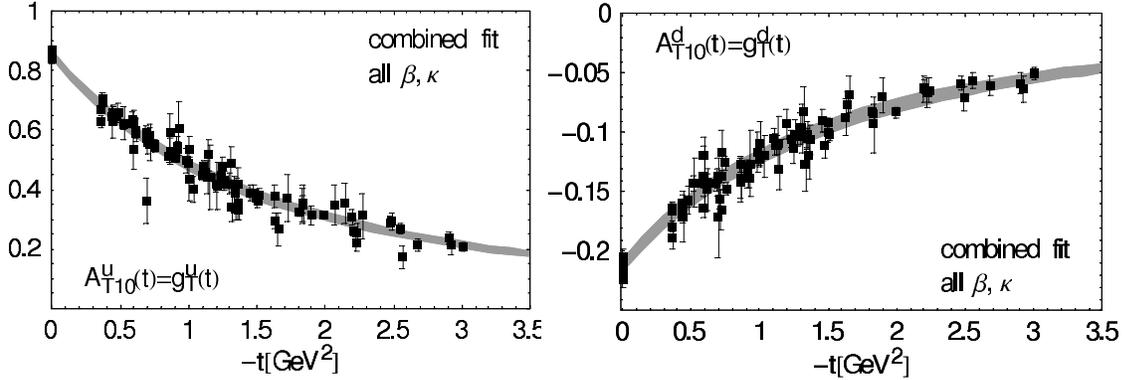}
\caption{The tensor form factor together with a dipole fit.}
\label{fig:AT10}
\ec
\end{figure}
\begin{figure}[t]
\bc
\includegraphics[height=15cm,angle=-90]{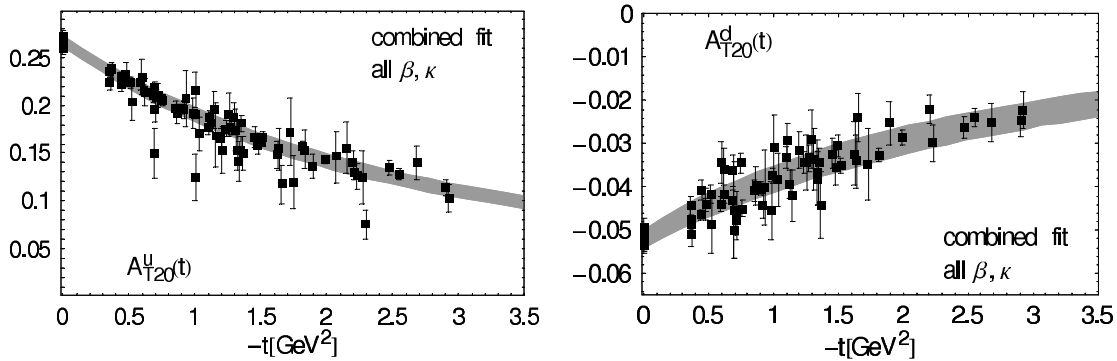}
\caption{The GFF $A_{T20}$ together with a dipole fit.}
\label{fig:AT20}
\ec
\end{figure}
Correlation functions are calculated on configurations taken at a
distance of 5-10 trajectories using 4-8 different locations of the
fermion source.
We use binning to obtain an effective distance of 20 trajectories. The
size of the bins has little effect on the error, which indicates
auto-correlations are small.
In this work, we simulate with three choices of sink momenta $\vec{P'}$ and
polarization operators, namely
\begin{equation}
\vec{P}_0' = ( 0, 0, 0 ),\,\,
\vec{P}_1' = ( \frac{2\pi}{L_S}, 0, 0 ),\,\,
\vec{P}_2'  = ( 0, \frac{2\pi}{L_S}, 0 )\ ,
\end{equation}
where $L_S$ is the spatial extent of the lattice, and
\begin{equation}
\tilde\Gamma_{\rm unpol} = \frac{1}{2}(1+\gamma_4),\,\,
\tilde\Gamma_1 = \frac{1}{2}(1+\gamma_4)\, i\gamma_5\gamma_1,\,\,
\tilde\Gamma_2 = \frac{1}{2}(1+\gamma_4)\, i\gamma_5\gamma_2 \ .
\label{projectors}
\end{equation}
The choice of the two polarization projectors, $\tilde\Gamma_1$ and
$\tilde\Gamma_2$ is particularly advantageous for the extraction of
the tensor GFFs.
The values of the momentum transfer $\vec{\Delta} = (2\pi/L_S)
\,\vec{q}$ used in this analysis are
\begin{equation}
\vec{q} : ( 0, 0, 0 ),\,\,( 1, 0, 0 ),\,\,( 1, 1, 0 )\,\,\,( 1, 1, 1 ),\,\,( 2, 0, 0 )
\end{equation}
and the vectors with permuted components.
\begin{figure}[t]
\bc
\includegraphics[height=15cm,angle=-90]{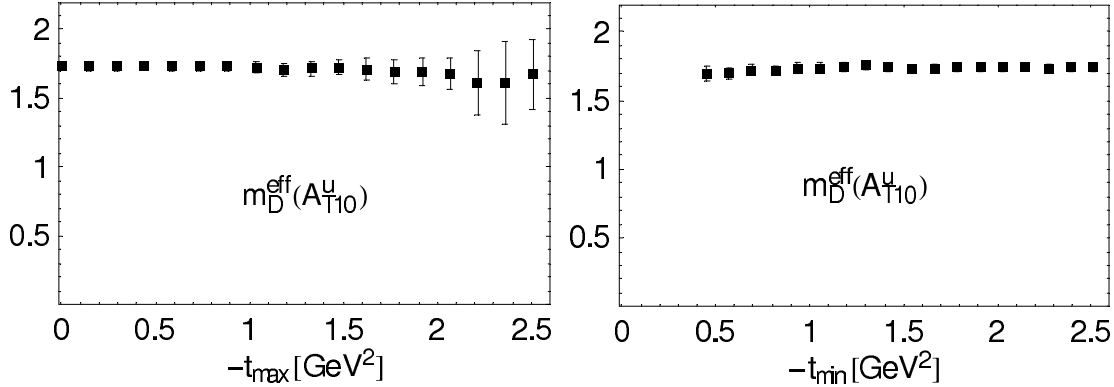}
\caption{The effective dipole mass as a function of a cut in $t$.}
\label{fig:DipoleCuts}
\ec
\end{figure}
All lattice results below have been non-perturbatively renormalized
\cite{reno} and transformed to the $\msbar$ scheme at a
renormalization scale of $4$ GeV$^2$.

In this work, we focus on the lowest two moments of the GPD $H_T$.  A
broader analysis will in particular include moments of the linear
combination $2\widetilde{H}_T(x,\xi,t) + E_T(x,\xi,t)$ which have been
shown to play a fundamental role for the transverse spin structure of
the nucleon \cite{Diehl:2005jf}. Furthermore, in \cite{Burkardt:2005hp} 
it is claimed that the $x$-moment of this linear combination gives the 
angular momentum carried by quarks with transverse spin
in an unpolarized nucleon, in analogy to Ji's sum rule.
%
%
In Figs.~\ref{fig:AT10} and \ref{fig:AT20} we show our results for the
lowest two moments of the generalized transversity for up and down
quarks in the nucleon as functions of the squared momentum transfer
$t$.
The lattice points and dipole curves are the result of a combined
dipole fit together with linear continuum and pion-mass extrapolations
of the form
\begin{equation}
A_{Tn0}^{\text{dipole},m_\pi,a} (t) = 
 \frac{A_{Tn0}^{0}(0)+\alpha_1 m_\pi^2+\alpha_2 a^2}
  {\left( 1 - {t/(m_D^0+\alpha_3 m_\pi^2)^2}
   \right)^2} \ ,
\label{chiralcontdipole}
\end{equation}
with five fit parameters $A_{Tn0}^{0}(0)$, $m_D^0$ and $\alpha_1,
\ldots, \alpha_3$. 
The curves show the fit function in the continuum limit, i.e.\ for
$a=0$, at the physical pion mass. Correspondingly, the difference
$A_{Tn0}^{\text{dipole},m_{\pi}^\text{latt},a}(t)-
A_{Tn0}^{\text{dipole},m_{\pi}^\text{phys},a=0}(t)$
has been subtracted from the individual data points before plotting.
Although the extrapolation to the continuum limit turns out to be
almost flat, except for $A^u_{20}(0)$ for which $\alpha_2\approx
-4.2\pm .7$ fm$^{-2}$, we include the $a^2$-dependence because it
reduces the $\chi^2$ of the fits considerably.
To check our ansatz in Eq.~(\ref{chiralcontdipole}), we show in
Fig.~\ref{fig:DipoleCuts} the (effective) dipole mass $m_D^0$ as a
function of a cut for minimal and maximal values of the momentum
transfer squared $t$ used for the fit, $t_\text{min}<t<t_\text{max}$
(keep in mind that $t<0$).
The effective dipole mass is in both cases very stable and constant,
except when $-t_\text{max}$ becomes large since there are not enough
data points used in the fit to determine the dipole mass accurately.
Still, a more sophisticated approach is desired for future
investigations.
Additionally, the assumed linear dependence on $a^2$ and $m_\pi^2$
eventually has to be replaced by a functional form obtained from e.g.~
chiral perturbation theory.
The quark mass dependence of the first two moments of the
(iso-vector) transversity has already been investigated in
\cite{Khan:2004vw,Detmold:2002nf}.

The forward moments and dipole masses at $m_\pi=m_{\pi}^\text{phys}$
and $a=0$ are found to be
\begin{equation}
\begin{array}{ll}
 \langle 1\rangle^u_\delta=A^u_{T10}(0) = 0.857\!\pm\! .013,  & m_D = 1.732 \!\pm\! .036 \text{ GeV}, \\
 \langle 1\rangle^d_\delta=A^d_{T10}(0) = -0.212\!\pm\! .005,  & m_D = 1.741 \!\pm\! .056 \text{ GeV}, \\
 \langle x\rangle^u_\delta=A^u_{T20}(0) = 0.268\!\pm\! .006,  & m_D = 2.312 \!\pm\! .071 \text{ GeV}, \\
 \langle x\rangle^d_\delta=A^d_{T20}(0) = -0.052\!\pm\! .002,  & m_D = 2.448 \!\pm\! .173 \text{ GeV}, \\
\label{res1}
\end{array}
\end{equation}
and for the iso-vector and iso-singlet combinations we obtain the
dipole masses
\begin{equation}
\begin{array}{ll}
 A_{T10}: & m_D^{u-d} = 1.731 \!\pm\! .034 \text{ GeV},\,\, 
            m_D^{u+d} = 1.713 \!\pm\! .043 \text{ GeV},\\
 A_{T20}: & m_D^{u-d} = 2.318 \!\pm\! .067 \text{ GeV},\,\, 
            m_D^{u+d} = 2.286 \!\pm\! .083 \text{ GeV},
\label{res2}
\end{array}
\end{equation}
which agree with the up- and down-quark dipole masses within errors.
Our result for the iso-vector tensor charge $\langle
1\rangle_{\delta}^{u-d}=1.068\pm0.016$ is in agreement with results in
\cite{Aoki:1996pi} and $5\%$ to $15\%$ lower compared to lattice
studies in
\cite{Gockeler:1997xy,Detmold:2002nf,Dolgov:2002zm,Orginos:2005uy}.
%
However our result for the iso-vector $x$-moment $\langle
x\rangle^{u-d}_\delta=0.322\pm0.006$ is substantially lower than the
quoted value of $\langle x\rangle^{u-d}_\delta=0.533\pm0.083$
(unquenched, $\kappa=0.1570$, from \cite{Dolgov:2002zm}) and also the
chirally extrapolated value $\langle
x\rangle^{u-d}_\delta=0.506\pm0.089$ \cite{Detmold:2002nf}
\footnote{This holds also for up and down quarks separately.} .
Since previous works used unimproved Wilson fermions with no
continuum extrapolation together with perturbative renormalization of
the operators, the numbers should be compared with some care.
Still, the discrepancy could indicate some problems with the normalization.
%
%

\begin{figure}[t]
\bc
\includegraphics[height=15cm,angle=-90]{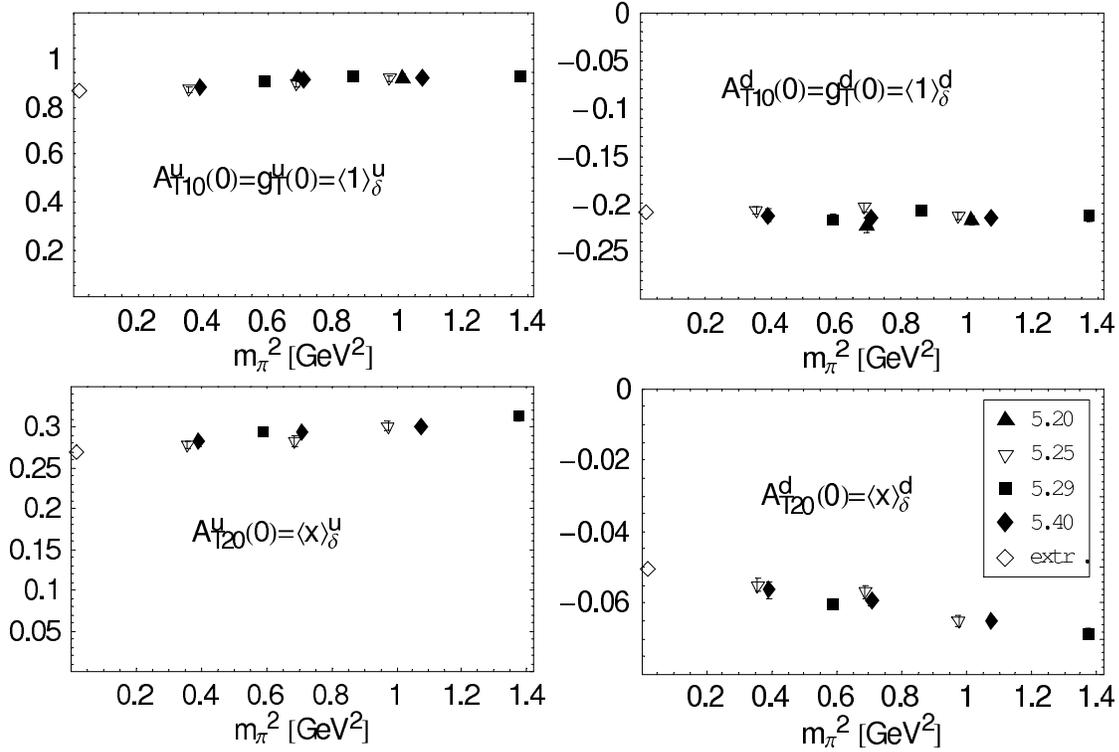}
\caption{The pion mass dependence of the lowest two moments
of the transversity distribution.}
\label{ATn0mPi}
\ec
\end{figure}

The explicit dependence of the tensor charge $g_{T}(t=0)=\langle 1
\rangle_\delta$ and the $x$-moment of the transversity $\langle x
\rangle_\delta$ on the pion mass is shown in Fig.~\ref{ATn0mPi}, where
all points have already been extrapolated to the continuum limit.  The
linearly extrapolated values at $m_{\pi}^\text{phys}$ agree within
errors with the results from the global fit in Eq.~(\ref{res1}).
From the figures we see that the tensor charge is approximately
constant over the available range of pion masses, while e.g. $\langle
x \rangle_\delta^d$ clearly shows a dependence on
$m_{\pi,\text{phys}}$ and drops by $\approx 20\%$ going from
$m_{\pi}^2=1.4$ GeV$^2$ down to $m_{\pi}^2=0.4$ GeV$^2$.

Interestingly, our results for the iso-vector dipole masses for the
first two moments of $H_T$ agree very well with those obtained from
fits to the moments of the polarized GPD, $\widetilde{H}$
\cite{GPDprep}, which are shown to lie on a linear Regge trajectory.
It will be interesting to see if this trend continues for higher
moments. 

\begin{figure}[t]
\bc
\includegraphics[height=15cm,angle=-90]{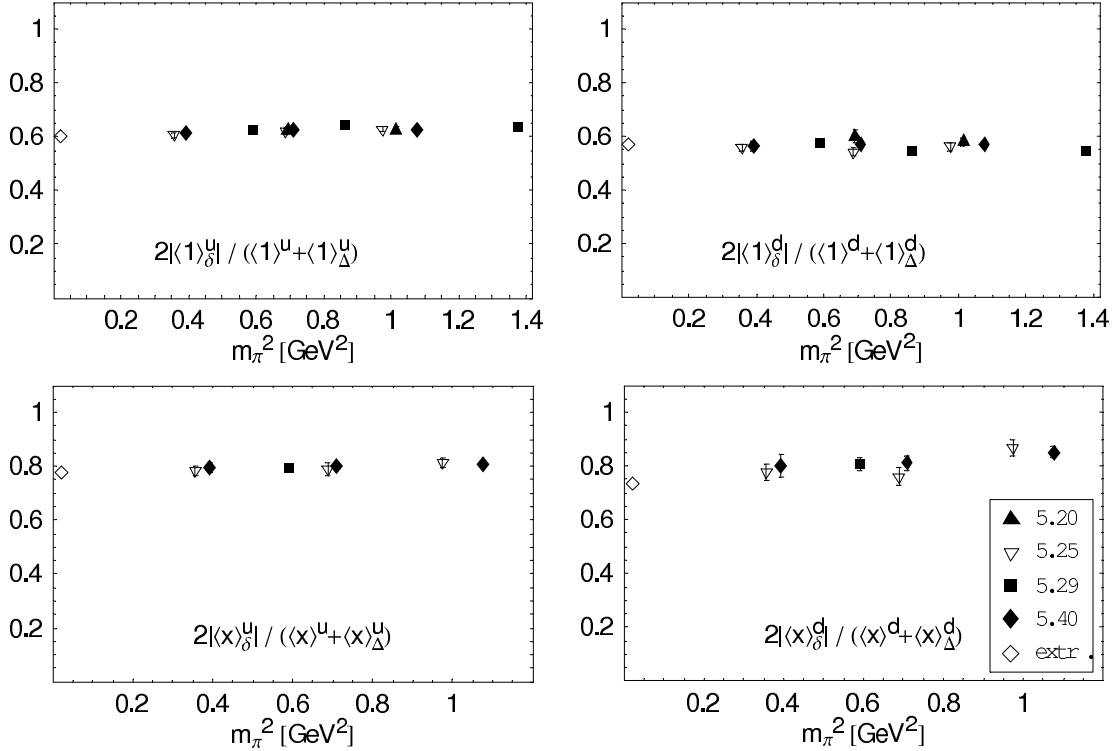}
\caption{The ratio in Eq.~(\ref{eq:sofferratio}) as a function of
  $m_\pi^2$ for $n=0,1$.}
\label{fig:soffer}
\ec
\end{figure}
Finally, in Fig.~\ref{fig:soffer} we investigate the Soffer bound
\cite{Soffer:1994ww}
\begin{equation}
\left|\delta q(x) \right|\le \frac{1}{2} \left( \Delta q(x) + q(x) \right)\ ,
\label{soffereq}
\end{equation}
which holds exactly only for quark and anti-quark distributions separately.
%
Mellin moments of the distribution functions as defined in section \ref{moments} give however
always sums/differences of moments of quark and anti-quark distributions,
e.g.~$\langle x^n\rangle_q+(-1)^{n+1}\langle x^n\rangle_{\bar q}$.
Taking Mellin moments of Eq.~(\ref{soffereq}) and
assuming that the antiquark contributions are small, we expect that the ratio
\begin{equation}
\frac{2\left|\langle x^n\rangle_{\delta}\right|}
     {\left( \langle x^n\rangle +
                        \langle x^n\rangle_{\Delta} \right)}\ ,
\qquad n=0,1\ ,
\label{eq:sofferratio}
\end{equation}
is smaller than one.
In Fig.~\ref{fig:soffer}, we show this ratio for up and down contributions as a function of $m_\pi^2$.
As we can clearly see from the figure, the ratio in Eq.~(\ref{eq:sofferratio}) is smaller than one
over the whole range of available pion masses. Taking into account what has been said above,
this strongly indicates that the Soffer bound is satisfied in our lattice calculation 
of the lowest two moments of the unpolarized, polarized and transversity quark distributions.


\vspace*{-3mm}
\section{Conclusions and Outlook}
\vspace*{-2mm}

We have computed the lowest moments of the quark tensor GPD $H_T$ in
lattice QCD and studied the chiral and continuum limit of the forward
moments and the dipole masses. Assuming that contributions from 
anti-quarks are small, our results indicate that the Soffer bound,
relating the transversity, unpolarized and polarized quark distributions,
is satisfied in our calculation.

The results are promising and our study will soon be extended to
include the tensor GPDs $E_T$, $\widetilde H_T$ and $\widetilde E_T$.
Once a set of the lowest moments of all tensor GPDs is available, it
will be extremely interesting to analyze the transverse spin density
of quarks in the nucleon, the corresponding positivity bounds and the
relation to moments of twist-3 GPDs using sum-rules obtained from the
equation of motion \cite{Diehl:2005jf}.


\vspace*{-3mm}
\section*{Acknowledgments}
\vspace*{-2mm}

The numerical calculations have been performed on the Hitachi SR8000
at LRZ (Munich), on the Cray T3E at EPCC (Edinburgh) under PPARC grant
PPA/G/S/1998/00777 \cite{Allton:2001sk}, and on the APEmille at NIC/DESY
(Zeuthen). This work is supported in part by the DFG (Forschergruppe
Gitter-Hadronen-Ph\"anomenologie), by the EU
Integrated Infrastructure Initiative Hadron Physics under contract
number RII3-CT-2004-506078 and by the Helmholtz Association, contract
number VH-NG-004.

\end{document}